\documentclass[12pt,reqno]{amsart}

\usepackage{amssymb,amsthm,amsmath,mathrsfs}
\usepackage{enumerate}
\usepackage{fullpage}

\usepackage{xcolor}

\begin{document}

\newtheorem{theorem}{Theorem}[section]
\newtheorem{lemma}[theorem]{Lemma}
\newtheorem{define}[theorem]{Definition}
\newtheorem{remark}[theorem]{Remark}
\newtheorem{corollary}[theorem]{Corollary}
\newtheorem{example}[theorem]{Example}
\newtheorem{assumption}[theorem]{Assumption}
\newtheorem{proposition}[theorem]{Proposition}
\newtheorem{conjecture}[theorem]{Conjecture}

\def\Ref#1{Ref.~\cite{#1}}

\def\Rnum{{\mathbb R}}
\def\const{\text{const.}}
\def\smallbinom#1#2{{\textstyle \binom{#1}{#2}}}
\def\smallsum{\textstyle\sum}

\def\pr{{\rm pr}}
\def\X{\mathbf{X}}
\def\Y{\mathbf{Y}}
\def\id{\mathrm{id}}
\def\lieder#1{{\mathcal L}_{#1}}

\def\Jsp{\mathrm{J}^{(\infty)}}
\def\Esp{\mathcal{E}}

\def\Rop{{\mathcal R}}
\def\Dop{{\mathcal D}}
\def\Hop{{\mathcal H}}
\def\Jop{{\mathcal J}}
\def\p{{\mathrm p}}
\def\q{{\mathrm q}}

\def\var{\text{var.}}
\def\dil{\text{dil.}}
\def\b{\beta}

\def\z{h}
\def\yone{A}
\def\ytwo{B}

\tolerance=50000
\allowdisplaybreaks[4]

\title{Symmetry analysis and hidden variational structure\\ of Westervelt's equation in nonlinear acoustics}

\author{
Stephen C. Anco$^1$,  
Almudena P. M\'arquez$^2$,\\
Tamara M. Garrido$^2$,
Mar\'ia L. Gandarias$^2$
\\\\
${}^1$D\lowercase{\scshape{epartment}} \lowercase{\scshape{of}} M\lowercase{\scshape{athematics and}} S\lowercase{\scshape{tatistics}}\\
B\lowercase{\scshape{rock}} U\lowercase{\scshape{niversity}}\\
S\lowercase{\scshape{t.}} C\lowercase{\scshape{atharines}}, ON L2S3A1, C\lowercase{\scshape{anada}} \\
\\
${}^2$D\lowercase{\scshape{epartment}} \lowercase{\scshape{of}} M\lowercase{\scshape{athematics}}\\
U\lowercase{\scshape{niversity of}} C\lowercase{\scshape{adiz}}\\
11510 P\lowercase{\scshape{uerto}} R\lowercase{\scshape{eal}}, C\lowercase{\scshape{adiz}}, S\lowercase{\scshape{pain}}\\
}


\begin{abstract}
Westervelt's equation is a nonlinear wave equation that is widely used to model 
the propagation of sound waves in a compressible medium, 
with one important application being ultra-sound in human tissue. 
Two fundamental aspects of the general dissipative version of Westervelt's equation
--- symmetries and conservation laws --- are studied 
in the present work by modern methods. 
Numerous results are obtained: 
new conserved integrals; 
potential systems yielding hidden symmetries and nonlocal conservation laws;
mapping of Westervelt's equation in the undamped case into a linear wave equation;
hidden variational structures, including a Lagrangian and a Hamiltonian;
a recursion operator and a Noether operator;
contact symmetries; higher-order symmetries and higher-order conservation laws. 
\end{abstract}

\maketitle

\section{Introduction}
Propagation of sound waves in a compressible medium \cite{HamBla-book}
has several important applications
where nonlinear and dissipative effects are relevant.
Examples are
(see e.g.\ \cite{Lur,GanYanKam,Cam,MyePylGilCamChiLog,MasLor,MaeSev,MenCaiLiZHoNiuZhe})
parametric arrays in water and in air,
under water imaging, 
musical acoustics of brass instruments, 
sonochemistry, 
quality control and characterization of materials,
and bio-medical devices.
Especially significant is ultra-sound imaging in human tissue
(see e.g.\ \cite{Sza,GueMarEglAliSer}). 

A simple mathematical 1D model is given by 
a dissipative version of Westervelt's equation \cite{Wes,Tar,Jor} 
\begin{equation}\label{model}
(1 -2\b p)p_{tt} - \alpha p_{ttt} - 2\beta p_t^2= c^2 p_{xx}
\end{equation}
describing the pressure fluctuation $p(t,x)$, 
where $\alpha>0$ is the damping coefficient
and $\b>0$ is the nonlinearity coefficient
which arises from the equation of state 
for the density $\rho(t,x)$ in terms of the pressure
\begin{equation}\label{eos}
\rho \approx p -\beta p^2 -\alpha p_t.
\end{equation}
For mathematical convenience, units will be chosen so that the sound speed
(in the linearized approximation) is $c=1$ hereafter. 
Note that equation \eqref{model}
can be written more succinctly as 
\begin{equation}\label{Weqn}
(p - \b p^2 - \alpha p_t)_{tt} = p_{xx}.
\end{equation}

Symmetries and conservation laws are intrinsic, fundamental aspects
of wave equations.
Their existence is not precluded by dissipative and nonlinear effects.
For instance, 
since $t$ and $x$ do not appear explicitly in the dissipative Westervelt equation \eqref{Weqn}, 
this gives rise to time-translation and space-translation symmetries,
which imply the existence of travelling waves; 
and since this equation has a second-order divergence form
in $t$ and $x$ derivatives,
it automatically possesses a conserved mass and a conserved center of mass,
holding for all solutions. 
Uncovering a richer structure of explicit symmetries and conservation laws 
in a given wave equation typically leads to numerous useful developments
concerning solutions and their properties. 

The present work is devoted to illustrating some of these developments
for the dissipative Westervelt equation \eqref{Weqn}, 
specifically:
\begin{itemize}
\item
new conserved integrals;
\item
potential systems yielding hidden symmetries and nonlocal conservation laws;
\item   
mapping of Westervelt's equation in the undamped case into a linear equation;
\item
hidden variational structures, including a Lagrangian and a Hamiltonian;
\item
a recursion operator and a Noether operator;
\item
contact symmetries; higher-order symmetries and higher-order conservation laws.
\end{itemize}
Symmetry multi-reduction and exact group-invariant solutions,
as well as mapping solutions to other solutions, 
which are additional main uses of symmetries, will be pursued in separate work.

In section~\ref{sec:symms.conslaws}, 
the Lie point symmetries of the dissipative Westervelt equation are determined. 
Since this equation does not have a local Lagrangian formulation
in terms of the given variable $p$, 
Noether's theorem is inapplicable 
and instead its modern generalization using multipliers 
is employed to determine the low-order conservation laws. 
These conservation laws yield five conserved integrals:
four of them are related to the net mass displaced by a sound wave; 
the fifth one turns out to be part of a hierarchy of higher order conservation laws,
which are not directly related to kinematic conserved quantities. 

In section~\ref{sec:potentialsys}, 
starting from the potential system arising via
the conserved form of the dissipative Westervelt equation, 
a second-layer potential is introduced. 
The Lie point symmetries and low-order conservation laws
for the second-layer potential system are derived. 
These results include two potential symmetries and three nonlocal conservation laws 
which are not inherited from any of the local point symmetries
and local conservation laws of the dissipative Westervelt equation. 
The conserved integrals given by the nonlocal conservation laws
are shown to describe net mass and a moment of net mass, 
as well as five additional quantities in the undamped case:
energy, momentum, dilational energy, dilational momentum,
and a generalized energy-momentum. 

In section~\ref{sec:results}, 
the main additional results, listed in the preceding bullet points,
are presented.
These results come from the second-layer potential system
and are lifted back to the dissipative Westervelt equation,
which yields a variety of hidden nonlocal structures.

Finally, in section~\ref{sec:conclude}, 
some concluding remarks are made. 

Throughout, we work in the setting of jet space,
using tools from variational calculus.
Notation and definitions are stated in Appendix~A. 
See \Ref{Ovs-book,Olv-book,BCA-book,Anc-review} for the basic theory of 
symmetries, multipliers, and variational structures for PDEs, 
presented in a form relevant for the present work.

All computations, including the proofs of the theorems,
have been done using Maple.
Some details are summarized in Appendix~B.

Recent work on the Westervelt equation in three spatial dimensions
has appeared in \Ref{SolSheThi,ManSolShe} on numerical studies,
\Ref{Kal} on analysis of the initial-value value problem,
and Lie point symmetries and group-invariant solutions in \Ref{Chi}.

\section{Point symmetries and low-order conservation laws}\label{sec:symms.conslaws}

A Lie point symmetry of the dissipative Westervelt equation \eqref{Weqn} 
is a group of point transformations under which the equation is invariant. 
The transformations have the infinitesimal form 
\begin{equation}\label{infinitesimal.point.transf}
t \to t +\epsilon \tau(t,x,p) + O(\epsilon^2), 
\quad
x \to x +\epsilon \xi(t,x,p) + O(\epsilon^2), 
\quad
p \to p +\epsilon \eta(t,x,p) + O(\epsilon^2), 
\end{equation}
acting on $(t,x,p)$ with $\epsilon\in\Rnum$ being the group parameter. 
Invariance holds if and only if  
\begin{equation}\label{inv.cond}
\big(\pr\X( p_{tt} - \b (p^2)_{tt} - \alpha p_{ttt} - p_{xx} )\big)\big|_\Esp =0
\end{equation}
where 
\begin{equation}\label{X.generator}
\X=\tau(t,x,p)\partial_t + \xi(t,x,p)\partial_x + \eta(t,x,p)\partial_p
\end{equation}
is the operator generating an infinitesimal point transformation \eqref{infinitesimal.point.transf}; 
$\pr\X$ denotes its prolongation;
and $\Esp$ denotes the space of solutions of equation \eqref{Weqn},
which is given by the equation and its differential consequences in the jet space. 

The invariance condition \eqref{inv.cond} constitutes 
a determining equation for symmetries.
In particular,
it splits with respect to derivatives of $p$ into
an overdetermined linear system 
which can be solved straightforwardly for
the functions for $\tau$, $\xi$, $\eta$. 
This system determines the admitted Lie point symmetries.

\begin{theorem}\label{thm:pointsymms}
The infinitesimal Lie point symmetries of the dissipative Westervelt equation \eqref{Weqn}
with $\b\neq 0$
are comprised by the linear span of 
a time-translation, a space-translation, a scaling combined with a shift
\begin{equation}\label{trans.scalshift.symm}
\X_1=\partial_t,
\quad
\X_2=\partial_x,
\quad
\X_3 = 2\b t \partial_t +3\b x\partial_x +(1-2\b p)\partial_p, 
\end{equation}
and in the undamped case, 
a dilation
\begin{equation}\label{dil.symm}
\X_4 = t\partial_t + x\partial_x, 
\quad
\alpha=0.
\end{equation}
\end{theorem}

Compared to the Lie point symmetries known in the dissipative case in three dimensions \cite{Chi},
the first three symmetries \eqref{trans.scalshift.symm} are inherited by reduction;
in the undamped case,
the fourth symmetry \eqref{dil.symm} also is inherited from an analogous symmetry
in three dimensions, 
although this case was not considered in \Ref{Chi}. 

The infinitesimal action of a Lie point symmetry on a solution $p(t,x)$
can be obtained by considering an equivalent generator,
called the characteristic (or evolutionary) form of $\X$, 
\begin{equation}\label{P.generator}
\hat\X=P\partial_p,
\quad
P = \eta(t,x,p) - \tau(t,x,p) p_t - \xi(t,x,p) p_x ,
\end{equation}
in which only $p$ undergoes a transformation. 
For the four symmetries \eqref{trans.scalshift.symm}--\eqref{dil.symm},
their characteristic form is given by 
\begin{equation}\label{P.symms}
\begin{aligned}
&
P_1 = -p_t ,
\quad
P_2 = -p_x,
\quad
P_3 = 1 - 2\b  p -2\b  t p_t -3\b  x p_x,
\\&
P_4 = -t p_t -x p_x,
\quad
\alpha =0 . 
\end{aligned}
\end{equation}
The symmetry determining equation \eqref{inv.cond} has an equivalent formulation
directly in terms of the function $P$:
\begin{equation}\label{symm.deteqn}
\big(\pr\hat\X( p_{tt} - \b (p^2)_{tt} - \alpha p_{ttt} - p_{xx} )\big)\big|_\Esp 
= \big( D_t^2 (P -2\b p P) -\alpha D_t^3 P  -  D_x^2 P \big)\big|_\Esp =0 
\end{equation}
following from the property that the prolonged generator $\pr\hat\X$ 
commutes with total derivatives $D$. 

A conservation law of the dissipative Westervelt equation \eqref{Weqn}
is a continuity equation
\begin{equation}\label{conslaw}
(D_t T + D_x \Phi)|_\Esp =0
\end{equation}
holding for all solutions $p(t,x$),
where $T$ is the conserved density and $\Phi$ is the spatial flux. 
Both $T$ and $\Phi$ are functions of $t$, $x$, $p$, and derivatives of $p$,
such that they are non-singular for all solutions.
The pair $(T,\Phi)$ is called a conserved current.

If $T=D_x\Theta$ and $\Phi=-D_t\Theta$ hold for all solutions,
where $\Theta$ is a function of $t$, $x$, $p$, and derivatives of $p$,
then the continuity equation holds identically and contains no useful
information about solutions $p(t,x)$.
Such a conservation law is called trivial.
If two conservation laws differ by trivial conservation law,
then they are said to be locally equivalent.
Hence,
only non-trivial conservation laws (up to local equivalence) are of interest.

Integration of a non-trivial conservation law over the spatial domain
$\Omega\subseteq \Rnum$ yields a conserved integral
\begin{equation}\label{cons.integral}
C=\int_{\Omega} T\,dx\big|_\Esp
\end{equation}
satisfying
\begin{equation}
\frac{dC}{dt} = -\Phi|_{\partial\Omega}\big|_\Esp.
\end{equation}
This states that the rate of change of the integral quantity \eqref{cons.integral}
is balanced by the net spatial flux leaving the domain $\Omega$
through the boundary points $\partial\Omega$. 
Under suitable boundary conditions posed on solutions $p(t,x)$,
the net flux will vanish, showing that $C$ is conserved
(namely, time-independent). 

When $\alpha\neq0$, the dissipative Westervelt equation is not of even order,
and hence it does not possess a Lagrangian formulation. 
Consequently, Noether's theorem is not applicable to find conservation laws.
Instead, all non-trivial conservation laws will arise from multipliers
as follows. 

A multiplier is a function of $t$, $x$, $p$, and derivatives of $p$
such that it is non-singular for all solutions and satisfies 
\begin{equation}\label{multreqn}
(p_{tt} - \b (p^2)_{tt} - \alpha p_{ttt} - p_{xx})Q= D_t T + D_x \Phi
\end{equation}
identically for some functions
$T$ and $\Phi$ of $t$, $x$, $p$, and derivatives of $p$.
In particular, since the dissipative Westervelt equation
is equivalent to an evolution system,
there is a one-to-one correspondence between
non-zero multipliers and non-trivial conserved currents (up to local equivalence),
after the highest-order $t$-derivative of $p$ in equation \eqref{Weqn}
is used to eliminate corresponding $t$-derivatives (and differential consequences)
in $Q$ and $(T,\Phi)$.
Then, from the multiplier equation \eqref{multreqn}, 
it is straightforward to show that
$Q=E_{p_t}(T)$ when $\alpha =0$, or $Q=E_{p_{tt}}(T)$ when $\alpha \neq 0$,
where $E_w$ denotes the Euler operator with respect to a variable $w$. 

A determining equation for multipliers is given by applying the Euler operator
with respect to $p$ to the multiplier equation \eqref{multreqn}:
\begin{equation}\label{multr.deteqn}
E_p( (p_{tt} - \b (p^2)_{tt} - \alpha p_{ttt} - p_{xx})Q ) =0
\end{equation}
which is required to hold identically and not just for solutions.
This Euler operator equation splits with respect to all derivatives of $p$
that do not appear in $Q$.
Hence, an overdetermined linear system for $Q$ is obtained,
which is similar to the overdetermined linear system for symmetries
in characteristic form given by $P$.
In particular, from general results in \Ref{Anc-review,AncBlu2002b}
for evolution systems, 
the multiplier system can be expressed as the adjoint of
the determining equation for symmetries, 
\begin{equation}\label{adjsymm.deteqn}
\big( (1 -2\b p) D_t^2 Q -\alpha D_t^3 Q - D_x^2 Q \big)\big|_\Esp =0 , 
\end{equation}
plus a set of Helmholtz-type equations.
Solutions of equation \eqref{adjsymm.deteqn} are called adjoint-symmetries. 

Typically, for wave equations,
conservation laws for basic physical quantities such as momentum and energy
come from multipliers of lower order than the order of the equation \cite{Anc-review}. 
Such low-order multipliers for equation \eqref{Weqn} would be of the form
$Q(t,x,p,p_t,p_x,p_{tt},p_{tx},p_{xx})$ when $\alpha\neq 0$,
and $Q(t,x,p,p_t,p_x)$ when $\alpha= 0$. 
It is straightforward to solve the overdetermined linear system
which determines these multipliers.

\begin{proposition}\label{prop:multr.loword}
The low-order multipliers for the dissipative Westervelt equation \eqref{Weqn}
with $\b\neq 0$
are comprised by the linear span of 
\begin{equation}\label{multr.loword}
Q_1 = 1,
\quad
Q_2= x,
\quad
Q_3 =t,
\quad
Q_4 =t\,x ,
\end{equation}
and in the undamped case, 
\begin{equation}\label{multr.loword.ais0}
Q_5 = p_t p_x/(p_x^2 -(1-2\b p) p_t^2)^2,
\quad
\alpha=0.
\end{equation}
\end{proposition}

The conserved current determined by a multiplier can be obtained
by several methods as explained in \Ref{Anc-review}.
This yields the following result.

\begin{theorem}\label{thm:conslaws}
The low-order conservation laws admitted by the dissipative Westervelt equation \eqref{Weqn}
consist of
\begin{align}
& 
T_1 = (1-2\b p)p_t -\alpha p_{tt}, 
&&
\Phi_1 = -p_x,
\label{conslaw1}\\
& 
T_2 = x((1-2\b p)p_t -\alpha p_{tt}), 
&&
\Phi_2 = p -x p_x,
\label{conslaw2}\\
&
T_3 = t((1-2\b p)p_t -\alpha p_{tt}) -(1-\b p)p +\alpha p_t, 
&&
\Phi_3 = -t p_x,
\label{conslaw3}\\
& 
T_4 =x(t((1-2\b p)p_t -\alpha p_{tt}) -(1-\b p)p +\alpha p_t), 
&&
\Phi_4 = t(p -x p_x),
\label{conslaw4}\\
& 
T_5 = p_x/(p_x^2 -(1-2\b p) p_t^2),
&&
\Phi_5 = p_t/(p_x^2 -(1-2\b p) p_t^2),
\quad
\alpha=0.
\label{conslaw5}
\end{align} 
\end{theorem}

The conserved quantities resulting from these conservation laws 
on the spatial domain $\Omega = (-\infty,\infty)$ 
will now be discussed.

\subsection{Conserved quantities}

Conservation laws \eqref{conslaw1} and \eqref{conslaw2} 
yield the conserved integrals
\begin{equation}\label{C1.C2}
C_1 = \int_{-\infty}^{\infty} ((1 -2\b p)p_t -\alpha p_{tt})\,dx, 
\quad
C_2 = \int_{-\infty}^{\infty} ((1 -2\b p)p_t - \alpha p_{tt})x\, dx.
\end{equation}
Their physical meaning is related to the mass displaced by a sound wave,
which can be seen when the integrals are written in terms of the density 
via the equation of state \eqref{eos}: 
\begin{equation}\label{C1.C2.mass.rel}
C_1\approx \int_{-\infty}^{\infty} \rho_t \,dx = \frac{d}{dt} m(t),
\quad
C_2\approx \int_{-\infty}^{\infty} \rho_t x\,dx = \frac{d}{dt} m^x(t)
\end{equation}  
where
\begin{equation}\label{mass.x-mass}
m(t) = \int_{-\infty}^{\infty} (\rho-\rho_0) \,dx,
\quad
m^x(t) = \int_{-\infty}^{\infty} (\rho-\rho_0) x\,dx,
\end{equation}    
with $\rho_0$ being the equilibrium (constant) density of the sound medium.
Here $m(t)$ is the net mass displaced by a sound wave,
where $\rho>\rho_0$ counts as a positive contribution,
while $\rho<\rho_0$ counts as a negative contribution. 
Similarly, $m^x(t)$ is the $x$-weighted net mass, 
which is proportional to the position of the center of mass defined by $m^x(t)/m(t)$. 
Therefore, 
conservation of $C_1$ implies $\frac{d^2}{dt^2}m(t)=0$, 
whereby $m(t)$ changes at a constant rate equal to $\frac{d}{dt}m(t) = C_1$, 
and hence $m(t)= m(0) + C_1 t$. 
Conservation of $C_2$ likewise implies $\frac{d^2}{dt^2}m^x(t)=0$, 
and thus $m^x(t) = m^x(0) + C_2 t$. 

Next, conservation laws \eqref{conslaw3} and \eqref{conslaw4} give rise to 
the integral quantities 
\begin{equation}\label{C3.C4.M.rel}
\begin{aligned}
C_3 & = \int_{-\infty}^{\infty} (T_3 +\rho_0)\,dx
\approx \int_{-\infty}^{\infty} (t\rho_t -(\rho-\rho_0))\,dx = t\frac{d}{dt} m(t) - m(t), 
\\
C_4 & = \int_{-\infty}^{\infty} (T_4 +x\rho_0)\,dx
\approx \int_{-\infty}^{\infty} (t\rho_t -(\rho-\rho_0))x\,dx = t\frac{d}{dt} m^x(t) - m^x(t).
\end{aligned}
\end{equation}
Time-independence of $C_3$ thereby implies 
$-C_3= -C_3|_{t=0} = m(0)$ 
which is the net mass initially displaced by a sound wave. 
Similarly, 
$-C_4= -C_4|_{t=0} = m^x(0)$ 
is the $x$-weighted net mass initially displaced by a sound wave. 
Thus, the initial position of the center of net mass of the wave is 
equal to $m^x(0)/m(0)=C_4/C_3$. 

Last, conservation law \eqref{conslaw5} yields
\begin{equation}\label{C5}
C_5 = \int_{-\infty}^{\infty} \frac{p_x}{p_x^2 +(2\b p-1) p_t^2} \,dx.
\end{equation}
This conserved integral has a distinctly different form 
compared to the previous quantities \eqref{C1.C2}--\eqref{C3.C4.M.rel}, 
and it does not have any apparent relationship to familiar kinematic quantities
such as mass, momentum, energy. 
Further discussion will be given in section~\ref{sec:higherorder}.

\section{Potential systems, symmetries and conservation laws}\label{sec:potentialsys}

Since the dissipative Westervelt equation \eqref{Weqn}
has the form of a continuity equation \eqref{conslaw},
given by the conserved current \eqref{conslaw1}, 
a potential $u(t,x)$ can be introduced such that
the equation becomes an identity via
\begin{subequations}\label{pot1.sys}
\begin{align}
(1-2\b p)p_t -\alpha p_{tt} & = u_x,
\label{pot1.eqn1}
\\
p_x & = u_t .
\label{pot1.eqn2}
\end{align}
\end{subequations}
Through equation \eqref{pot1.eqn2},
a second-layer potential $v(t,x)$ can be introduced: 
\begin{equation}\label{pot2.sys}
p =  v_t,
\quad
u = v_x.
\end{equation}
Then equation \eqref{pot1.eqn1} yields a potential equation
\begin{equation}\label{pot2.eqn}
(v_t -\b v_t^2 -\alpha v_{tt})_t = v_{xx}.
\end{equation}

Every solution $v(t,x)$ of the potential equation
yields a solution $p(t,x)=v_t(t,x)$ of the dissipative Westervelt equation
as shown by the relation 
\begin{equation}\label{Weqn.pot2eqn.rel}
p_{tt} - \b (p^2)_{tt} - \alpha p_{ttt} - p_{xx}
= D_t((v_t -\b v_t^2 -\alpha v_{tt})_t - v_{xx}).
\end{equation}

It is useful to note that the potential equation can be written
in a slightly simpler form
\begin{equation}\label{pot2alt.eqn}
(\b \tilde v_t^2 +\alpha \tilde v_{tt})_t = -\tilde v_{xx},
\quad
\tilde v = v - t/(2\b).
\end{equation}

\subsection{Potential point symmetries}
Symmetries (in characteristic form) $\hat\X^v=P^v\partial_v$ of the potential equation \eqref{pot2.eqn}
are determined by the equation
\begin{equation}\label{pot2.eqn.symm.deteqns}
( D_t( (1 -2\b v_t)D_t P^v ) -\alpha D_t^3 P^v  -D_x^2 P^v )|_{\Esp^v} =0,  
\end{equation}
where $\Esp^v$ denotes the solution space of equation \eqref{pot2.eqn},
which is given by the equation and its differential consequences in the jet space. 

This determining equation \eqref{pot2.eqn.symm.deteqns}
can be straightforwardly solved to obtain all Lie point symmetries 
of the potential equation \eqref{pot2.eqn},
with
\begin{equation}\label{Pv.generator}
P^v = \eta^v(t,x,v) - \tau^v(t,x,v) v_t - \xi^v(t,x,v) v_x
\end{equation}
being their characteristic form. 

\begin{theorem}\label{thm:pot.pointsymms}
The infinitesimal Lie point symmetries of potential equation \eqref{pot2.eqn}
with $\b\neq 0$
are comprised by the linear span of two shifts,
a time-translation, a space-translation, a scaling-shift,
and a scaling in the undamped case. 
Their respective characteristic forms are given by
\begin{equation}\label{pot2.eqn.symms1}
P^v_1 = 1,
\quad
P^v_2= x,
\quad
P^v_3= -v_t,
\quad
P^v_4 = -v_x,
\quad
P^v_5 = t -2\b t v_t -3\b x v_x, 
\end{equation}
and
\begin{equation}\label{pot2.eqn.symms2}
P^v_6 = v -t v_t -x v_x, 
\quad
\alpha=0.
\end{equation}
The latter four symmetries are inherited from the dissipative Westervelt equation \eqref{Weqn}
via the projection 
\begin{equation}\label{P.rel}
P=D_t P^v 
\end{equation}
which arises directly from $p=v_t$.
The two shifts are ``hidden'' symmetries which exist only for the potential equation. 
\end{theorem}

The specific correspondence between
the inherited Lie point symmetries of the potential equation
and the Lie point symmetries of equation \eqref{Weqn} is given by
\begin{equation}\label{P.Pv}
P^v_3 \leftrightarrow P_1;
\quad
P^v_4 \leftrightarrow P_2;
\quad
P^v_5 \leftrightarrow P_3;
\quad
P^v_6 \leftrightarrow P_4. 
\end{equation}

\subsection{Potential conservation laws}

Similarly to the situation for the dissipative Westervelt equation,
there is a one-to-one correspondence between
non-trivial conserved currents (up to local equivalence) $(T^v,\Phi^v)$
and non-zero multipliers $Q^v$
for the potential equation \eqref{pot2.eqn},
where 
\begin{equation}\label{pot2.multreqn}
((v_t -\b v_t^2 -\alpha v_{tt})_t - v_{xx})Q^v = D_t T^v + D_x \Phi^v.
\end{equation}
This correspondence holds if the highest-order $t$-derivative of $v$
in the potential equation is used to eliminate corresponding $t$-derivatives (and differential consequences) 
in $(T^v,\Phi^v)$ and $Q^v$,
whereby $Q^v=E_{v_t}(T)$ when $\alpha =0$, or $Q^v=E_{v_{tt}}(T)$ when $\alpha \neq 0$. 

The determining equation for multipliers $Q^v$ 
is given by the Euler operator equation
\begin{equation}\label{pot2.multr.deteqn}
E_v( ((v_t -\b v_t^2 -\alpha v_{tt})_t - v_{xx})Q^v ) =0
\end{equation}
which is required to hold identically. 
This equation splits with respect to all derivatives of $v$
that do not appear in $Q^v$, yielding an overdetermined linear system. 
The following result gives the solution of the system
for all low-order multipliers characterized by the general form
$Q(t,x,v,v_t,v_x,v_{tt},v_{tx},v_{xx})$ when $\alpha\neq 0$,
and $Q(t,x,v,v_t,v_x)$ when $\alpha= 0$. 

\begin{proposition}\label{prop:pot.multr.loword}
The low-order multipliers for potential equation \eqref{pot2.eqn} 
with $\b\neq 0$
are comprised by the linear span of 
\begin{equation}\label{pot2.multrs1}
Q^v_1 = 1,
\quad
Q^v_2 = x,
\end{equation}
and in the undamped case, 
\begin{align}
& Q^v_3 =\tfrac{2}{\b}t + v -5 t v_t - 7x v_x, 
\label{pot2.multr2}\\
& Q^v_4 =
v v_x + t (\tfrac{2}{\b}  - 5v_t)v_x -x (4v_x^2 +\tfrac{1}{3\b^2}(1-2\b v_t)^3),
\label{pot2.multr3}\\
& Q^v_5 = f(v_t,v_x),
\label{pot2.multr4.funct}
\end{align}
where
\begin{equation}\label{pot2.f.eqn}
f_{v_t v_t} = (1-2\b v_t) f_{v_x v_x}.
\end{equation}
\end{proposition}

Use of any of the methods explained in \Ref{Anc-review} yields 
the conserved currents arising from these multipliers. 
The simplest form for them is obtained by working 
in terms of the variable \eqref{pot2alt.eqn},
which gives the following result. 

\begin{theorem}\label{thm:pot2.conslaws}
The low-order conservation laws admitted by the potential equation \eqref{pot2.eqn}
consist of 
\begin{align}
& T^v_1 = \tfrac{1}{\b} (\b v_t -\tfrac{1}{2})^2 +\alpha v_{tt} , 
&&
\Phi^v_1 = v_x , 
\label{pot2.conslaw1}\\
& T^v_2 = \tfrac{1}{\b} x((\b v_t -\tfrac{1}{2})^2 + \alpha v_{tt}), 
&&
\Phi^v_2 = x v_x -v , 
\label{pot2.conslaw2}  
\end{align}
and in the undamped case,
\begin{align}
&\begin{aligned}
T^v_3 &= t (\tfrac{10}{3\b^2} (\b v_t -\tfrac{1}{2})^3 - \tfrac{5}{2}v_x^2)
+ \tfrac{7}{\b} x (\b v_t -\tfrac{1}{2})^2 v_x
- \tfrac{1}{\b} (\b v_t -\tfrac{1}{2})^2 (v -\tfrac{1}{2\b} t), 
\\
\Phi^v_3 &=
\tfrac{5}{\b} t (\b v_t -\tfrac{1}{2}) v_x + x (\tfrac{7}{2} v_x^2 -\tfrac{7}{3\b^2} (\b v_t -\tfrac{1}{2})^3) - v_x (v - \tfrac{1}{2\b}t), 
\end{aligned}
\label{pot2.conslaw3}\\
&\begin{aligned}
T^v_4 &= t (\tfrac{10}{3\b^2}(\b v_t -\tfrac{1}{2})^3 v_x - \tfrac{5}{6} v_x^3)
+ \tfrac{4}{\b} x (\b v_t -\tfrac{1}{2})^2 (v_x^2 -\tfrac{4}{15\b^2} (\b v_t -\tfrac{1}{2})^3)
\\&\quad
- \tfrac{1}{\b}(\b v_t -\tfrac{1}{2})^2 v_x (v-\tfrac{1}{2\b}t), 
\\
\Phi^v_4&=
t (\tfrac{5}{2\b}(\b v_t -\tfrac{1}{2})v_x^2  -\tfrac{5}{6\b^3} (\b v_t -\tfrac{1}{2})^4)
+ x (\tfrac{4}{3} v_x^3 -\tfrac{8}{3\b^2} (\b v_t -\tfrac{1}{2})^3 v_x)
\\&\quad
+ (\tfrac{1}{3\b^2} (\b v_t -\tfrac{1}{2})^3 -\tfrac{1}{2} v_x^2)(v -\tfrac{1}{2\b}t), 
\end{aligned}
\label{pot2.conslaw4}\\
&\begin{aligned}
T^v_5 &= \int  (1 -2\b v_t) f\, dv_t,
 \qquad
\Phi^v_5 &= \int (1 -2\b v_t) v_t f_{v_x}\, dv_t
- v_t \int (1 -2\b v_t) f_{v_x}\, dv_t, 
\end{aligned}
\label{pot2.conslaw5}
\end{align}
where $f(v_t,v_x)$ is an arbitrary solution of the linear PDE \eqref{pot2.f.eqn}. 
\end{theorem}

In the special cases $f=v_t-\tfrac{1}{2\b}$ and $f=v_x$,
which correspond to the multipliers
\begin{equation}\label{pot2.multr.funct.spec}
Q^v_{5a} = v_t-\tfrac{1}{2\b},
\quad
Q^v_{5b} = v_x, 
\end{equation}
the conservation law \eqref{pot2.conslaw5} can be simplified to the form
\begin{align}
& T^v_{5a} = \tfrac{1}{2} v_x^2 -\tfrac{2}{3\b^2} (\b v_t -\tfrac{1}{2})^3, 
&& \Phi^v_{5a} =-(v_t -\tfrac{1}{2\b})v_x, 
\label{pot2.conslaw5a}\\
& T^v_{5b} = \tfrac{1}{\b^2}(\b v_t-\tfrac{1}{2})^2 v_x ,
&& \Phi^v_{5b} = \tfrac{1}{2} v_x^2 -\tfrac{1}{3\b^2} (\b v_t -\tfrac{1}{2})^3, 
\label{pot2.conslaw5b}
\end{align}
respectively. 

The physical meaning of these conservation laws will be discussed in the next subsection. 

It is useful to observe that, in general, 
multipliers for the potential equation are related to
multipliers for the dissipative Westervelt equation by 
\begin{equation}\label{Q.rel}
Q^v=-D_t Q.
\end{equation}
To derive this relation,
consider the multiplier identity
\begin{equation}
D_t((v_t -\b v_t^2 -\alpha v_{tt})_t - v_{xx}) Q
= ((v_t -\b v_t^2 -\alpha v_{tt})_t - v_{xx})(-D_t Q)
+ D_t( ((v_t -\b v_t^2 -\alpha v_{tt})_t - v_{xx})Q )
\end{equation}
which is obtained from integration by parts applied to
the relation \eqref{Weqn.pot2eqn.rel} multiplied by $Q$.
The left-hand side of the identity will be a total divergence
when $Q$ is a multiplier for the dissipative Westervelt equation.
Applying the Euler operator $E_v$ then annihilates
all of the total derivative terms on both sides,
which yields the multiplier determining equation \eqref{pot2.multr.deteqn}
where $Q^v$ is given by the relation \eqref{Q.rel}. 

It is readily seen that the multipliers \eqref{pot2.multrs1} are inherited
from local multipliers for the dissipative Westervelt equation
\begin{equation}\label{Qv1.Qv2}
-D_t(Q_3) = -Q^v_1, 
\quad
-D_t(Q_4)= -Q^v_2 ,
\end{equation}
whereas the multipliers \eqref{pot2.multr2}--\eqref{pot2.multr4.funct}
correspond to nonlocal multipliers for the dissipative Westervelt equation.

\subsection{Nonlocal conserved quantities}

Every conservation law admitted by the potential equation \eqref{pot2.eqn}
holds for solutions of the dissipative Westervelt equation
due to the relation \eqref{Weqn.pot2eqn.rel}.

A potential conservation law will be a local conservation law of the dissipative Westervelt equation
if its conserved current $(T^v,\Phi^v)|_{\Esp^v}$
has no essential dependence on $v$ and $x$-derivatives of $v$,
up to the addition of a trivial current.

The potential conservation laws \eqref{pot2.conslaw1} and \eqref{pot2.conslaw2} 
have local conserved densities which coincide with some of the terms
in the densities of the respective local conservation laws \eqref{conslaw1} and \eqref{conslaw2}. 
It is straightforward to see, using the potential equation expressed as
$v_{xx} = (1-2\b p)p_t -\alpha p_{tt}$,
that $(T^v_1 -T^3)|_{\Esp^v} = D_x\Theta$ and $(\Phi^v_1 -\Phi^3)|_{\Esp^v} = -D_t\Theta$
holds for $\Theta = tv_x +\tfrac{1}{4\b} x$.
Hence, the conservation laws \eqref{pot2.conslaw1} and \eqref{conslaw1}
are locally equivalent.
Likewise, the conservation laws \eqref{pot2.conslaw2} and \eqref{conslaw2}
can be seen to be locally equivalent. 
The conserved quantities can thus be expressed entirely in terms of $p$: 
\begin{equation}\label{Cv1.Cv2}
C^v_1 = \int_{-\infty}^{\infty} ( (\b p -1)p +\alpha p_t)\, dx,
\quad
C^v_2 = \int_{-\infty}^{\infty} ( (\b p -1)p +\alpha p_t) x\, dx.
\end{equation}
Up to the addition of a constant,
the densities in these two conserved quantities
are the same as the mass density and the $x$-weighted mass density
appearing in the integrals \eqref{mass.x-mass},
as seen via the equation of state \eqref{eos}. 
Therefore, the net mass and the $x$-weighted mass are actually 
conserved quantities themselves. 

All of the other potential conservation laws \eqref{pot2.conslaw3}--\eqref{pot2.conslaw5} and \eqref{pot2.conslaw5a}--\eqref{pot2.conslaw5b}
have a nonlocal conserved density involving 
$v=\partial_t^{-1}p$ or $v_x=\partial_t^{-1}p_x$. 
The resulting conserved quantities (after scaling by a numerical factor) consist of 
energy 
\begin{equation}\label{ener}
E = \int_{-\infty}^{\infty} ( \tfrac{1}{2} v_x^2 -\tfrac{2}{3}\b (p -\tfrac{1}{2\b})^3 )\, dx , 
\end{equation}
momentum
\begin{equation}\label{mom}
M = \int_{-\infty}^{\infty} (p -\tfrac{1}{2\b})^2 v_x \, dx , 
\end{equation}
dilation-type energy
\begin{equation}\label{dilener}
K = \int_{-\infty}^{\infty}\big( 
t (\tfrac{1}{2}v_x^2 - \tfrac{2}{3}\b (p -\tfrac{1}{2\b})^3)
-\tfrac{1}{5}\b(7 x v_x -v +\tfrac{1}{2\b} t)(p -\tfrac{1}{2\b})^2 
\big)\, dx , 
\end{equation}
and dilation-type momentum, 
\begin{equation}\label{dilmom}
\begin{aligned}
H = \int_{-\infty}^{\infty}\big(
& \tfrac{1}{2} t(v_x^2 - 4 \b(p -\tfrac{1}{2\b})^3)
-\tfrac{3}{5} \b (4 x v_x -v + \tfrac{1}{2\b}t) (p -\tfrac{1}{2\b})^2 
+ (\tfrac{4}{5}\b)^2 x v_x^4
\big)v_x \, dx , 
\end{aligned}
\end{equation}
which respectively arise from the conservation laws \eqref{pot2.conslaw5a}, \eqref{pot2.conslaw5b}, \eqref{pot2.conslaw3}, and \eqref{pot2.conslaw4}. 
A generalized energy-momentum 
\begin{equation}\label{gen.enermom}
I = \int_{-\infty}^{\infty} F(p,v_x)\, dx,
\quad
F(p,v_x) = \int (1 -2\b p) f(p,v_x)\, dp
\end{equation}
arises from conservation law \eqref{pot2.conslaw5}. 

The physical meaning of the quantities \eqref{ener} and \eqref{mom}
can be seen by examining their densities when $\b$ is small,
whereby the potential equation \eqref{pot2.eqn} reduces to a linear equation
\begin{equation}\label{lin.pot2.eqn}
v_{tt} - \alpha v_{ttt} = v_{xx}.
\end{equation}
Consider, firstly, the momentum density $p v_x -\b p^2v_x -\tfrac{1}{4\b} v_x$. 
The last term is locally trivial,
$-\tfrac{1}{4\b} v_x = D_x\Theta$, with $\Theta = -\tfrac{1}{4\b} v$. 
Modulo this density, 
the remaining terms have the form $p v_x + O(\b)$ 
which reduces to the well-known momentum density $p v_x$ 
for the linear equation \eqref{lin.pot2.eqn}. 
Secondly, consider the energy density
$\tfrac{1}{2} v_x^2 +p^2 -\tfrac{2}{3}\b p^3 - \tfrac{1}{2\b} p - \tfrac{1}{12\b^2}$. 
The linear term in $p$ can be cancelled by 
adding a multiple of the density \eqref{pot2.conslaw1} 
specialized to the case $\alpha=0$:
$\b p^2 -p +\tfrac{1}{4\b}$. 
This cancellation corresponds to removing the constant term in the multiplier $Q^v_{5a}$,
so that the modified multiplier is simply $Q^v=v_t=p$. 
The resulting modified density has the form 
$\tfrac{1}{2} v_x^2 + \tfrac{1}{2} p^2 + O(\b) + D_x\Theta$ 
with $\Theta= - \tfrac{1}{24\b^2} x$. 
Hence, modulo a locally trivial density, 
the remaining terms reduce to the well-known energy density 
$\tfrac{1}{2} (v_x^2 +p^2)$ for the linear equation \eqref{lin.pot2.eqn}. 

The preceding argument does not work for the quantities \eqref{dilener} and \eqref{dilmom}
because they contain non-trivial terms involving inverse powers of $\b$. 
Their physical meaning as dilational quantities comes from a comparison of 
the form of the terms containing $t$
and the form of the terms in the energy and momentum densities. 
Specifically, for the density in the quantity \eqref{dilener}, 
the terms $t (\tfrac{1}{2}v_x^2 - \tfrac{2}{3}\b (p -\tfrac{1}{2\b})^3)$
are exactly $t$ times the density in the energy \eqref{ener}. 
In the quantity \eqref{dilmom}, the terms 
$\tfrac{1}{2} t(v_x^2 - 4 \b(p -\tfrac{1}{2\b})^3)v_x$ 
share the feature with the density in the momentum \eqref{mom}
that they are odd in $v_x$.

\section{Main results}\label{sec:results}

\subsection{Variational structure}\label{sec:variational}

The potential equation \eqref{pot2.eqn} has the property that
the symmetry determining equation \eqref{pot2.eqn.symm.deteqns}
in the undamped case, $\alpha=0$, is self-adjoint. 
Namely, the linear operator (in total derivatives)
$D_t(1 -2\b v_t)D_t -D_x^2$, 
which is the Frechet derivative of the potential equation with $\alpha=0$,
is equal to its adjoint as defined via integration by parts. 
Self-adjointness is well known to be the necessary and sufficient condition
for a given equation to be an Euler-Lagrange equation. 
Hence, the undamped potential equation
\begin{equation}\label{pot2.eqn.undamped}
(v_t -\b v_t^2)_t = v_{xx}
\end{equation}
has a Lagrangian formulation 
\begin{equation}\label{pot2.ELeqn}
G^v =(1 -2\b v_t)v_{tt} -v_{xx} = E_v(L)
\end{equation}
where the Lagrangian is straightforwardly found to be
\begin{equation}\label{pot2.Lagr}
L = \tfrac{1}{2} (v_x^2 -v_t^2)  +\tfrac{1}{3}\b v_t^3.
\end{equation}
Note that $L$ is unique only up to the addition of
an arbitrary total divergence.

An infinitesimal variational symmetry is a generator $\hat\X^v=P^v_\var\partial_v$
under which $L$ is invariant up to a total divergence, 
$\pr\hat\X^v(L) = D_t A + D_x B$,
for some functions $A$ and $B$ depending on $t$, $x$, $v$ and its derivatives.
This invariance implies that the extremals of $L$ are preserved
and hence $\hat\X^v=P^v_\var\partial_v$ will be an infinitesimal symmetry of
the undamped potential equation $G^v=0$.

Variational symmetries coincide with multipliers.
This is a consequence of the variational identity
$\pr\hat\X^v(L) = P^v_\var E_v(L) + D_t\Theta^t + D_x\Theta^x$ 
which yields
$P^v_\var E_v(L) = D_t(A-\Theta^t) + D_x(B-\Theta^x)$ 
from which Noether's theorem is obtained.
The latter equation is exactly the same as the multiplier equation \eqref{pot2.multreqn},
with the identification
\begin{equation}\label{pot2.PisQ}
P^v_\var = Q^v.
\end{equation}
Thus, the multiplier determining equation \eqref{pot2.multr.deteqn}
provides a determining equation for variational symmetries,
\begin{equation}\label{pot2.varsymm.deteqn}
E_v( ((v_t -\b v_t^2)_t - v_{xx})P^v_\var ) =0, 
\end{equation}
without the explicit use of $L$. 

Comparison of the point symmetries \eqref{pot2.eqn.symms1}--\eqref{pot2.eqn.symms2}
and the low-order multipliers \eqref{pot2.multrs1}, \eqref{pot2.multr2}, \eqref{pot2.multr.funct.spec}
shows that
\begin{equation}\label{pot2.QP}
Q^v_1 = P^v_1,
\quad
Q^v_2 = P^v_2,
\quad
Q^v_3 = P^v_6 +\tfrac{2}{\b} P^v_5,
\quad
Q^v_{5a} = -P^v_3,
\quad
Q^v_{5b} = -P^v_4
\end{equation}
represent variational Lie point symmetries.
The remaining multipliers \eqref{pot2.multr3} and \eqref{pot2.multr4.funct},
which are nonlinear in $v_t$ and $v_x$, 
represent first-order variational symmetries
\begin{equation}\label{pot2.contactP}
P^v_7 := Q^v_4,
\quad
P^v_8 := Q^v_5, 
\end{equation}
each of which can be expressed equivalently
as a contact symmetry \cite{BA-book}. 
Specifically, their respective canonical forms are given by 
\begin{equation}\label{pot2.contactsymm}
\begin{aligned}
\X^v & =
(5 t v_x - (2/\b) x (1 -2 \b v_t)^2)\partial_t
+ (8 x v_x +  t (5 v_t - 2/\b) - v)\partial_x
\\&\quad
+(4 x(v_x^2 +\tfrac{1}{3} (3 -4\b v_t)v_t^2) + 5 t v_x v_t)\partial_v
-2 (2 v_t - 1/\b)v_x \partial_{v_t}
\\&\quad
+ (-3 v_x^2 + (\tfrac{8}{3}\b v_t^2 - 4 v_t + 2/\b) v_t)\partial_{v_x}
\end{aligned}
\end{equation}
and
\begin{equation}\label{pot2.contactsymm.funct}
\X^v = -f_{v_t}\partial_t -f_{v_x}\partial_x + (f -v_t f_{v_t} - v_x f_{v_x})\partial_v 
\end{equation}
where $f(v_t,v_x)$ is an arbitrary solution of the linear PDE \eqref{pot2.f.eqn}.

\subsection{Hamiltonian formulation}\label{sec:hamil}

The Lagrangian structure \eqref{pot2.ELeqn} of
the undamped potential equation \eqref{pot2.eqn.undamped}
has a straightforward corresponding Hamiltonian formulation.

The Hamiltonian variables consist of
\begin{equation}
\q:= v,
\quad
\p:= \partial_{v_t}L = -v_t +\b v_t^2 = (\b p -1)p .
\end{equation}
A Legendre transformation applied to the Lagrangian \eqref{pot2.Lagr} yields 
$\p v_t - L = {-}\tfrac{1}{2} (v_x^2 +v_t^2)  +\tfrac{2}{3}\b v_t^3$,
which can be seen to be the negative of the density of the energy conserved integral \eqref{ener}. 
Taking the Hamiltonian to be energy then leads to the equations of motion
\begin{equation}\label{pot2.Hamil.eom}
\begin{pmatrix} v_t \\ \p_t \end{pmatrix}
= \Hop \begin{pmatrix} \delta E/\delta v \\ \delta E/\delta \p \end{pmatrix},
\quad
\Hop = \begin{pmatrix} 0 & -1\\ 1 & 0 \end{pmatrix}
\end{equation}
where 
$\delta E/\delta v = -v_{xx}$,
and $\delta E/\delta \p = -p$
which follows from 
$\delta E/\delta p = p-2\b p^2 = (\partial_p \p)\delta E/\delta \p$
and $\partial_p\p = 2\b p -1$. 
These equations of motion \eqref{pot2.Hamil.eom} yield
$v_t =p$ and $\p_t = -v_{xx}$,
which are equivalent to the undamped potential equation \eqref{pot2.eqn.undamped}.

\subsection{Noether operator}\label{sec:noetherop}

The variational structure \eqref{pot2.PisQ} 
can be lifted to the dissipative Westervelt equation \eqref{Weqn}
in the undamped case, $\alpha=0$, by use of relations \eqref{P.rel}, \eqref{Q.rel}, and \eqref{pot2.PisQ}. 
This yields
\begin{equation}\label{rels}
\begin{aligned}
Q^v & =-D_t Q
\\& = P^v_\var = D_t^{-1}P
\end{aligned}
\end{equation}
and hence 
\begin{equation}\label{noether.rel}
P=-D_t^2 Q.
\end{equation}
Consequently, multipliers of the undamped Westervelt equation
\begin{equation}\label{Weqn.undamped}
(p - \b p^2)_{tt} = p_{xx}
\end{equation}
are mapped into infinitesimal symmetries through 
\begin{equation}\label{invNoether.op}
\Jop^{-1} = -D_t^2
\end{equation}
which defines an inverse Noether operator.

Applying this operator \eqref{invNoether.op} to the low-order multipliers \eqref{multr.loword} and \eqref{multr.loword.ais0} 
yields, respectively, a trivial infinitesimal symmetry
and a third-order infinitesimal symmetry.
More remarkably,
the latter turns out to belong to a hierarchy of higher-order symmetries,
which will be derived later from a recursion operator applied to
the Lie point symmetries. 

The inverse of the operator \eqref{invNoether.op},
which constitutes a Noether operator 
\begin{equation}\label{Noether.op}
\Jop = -(D_t^{-1})^2 , 
\end{equation}
maps a subspace of infinitesimal symmetries into multipliers.
The subspace domain of this operator is defined by
the variational symmetry condition \eqref{pot2.varsymm.deteqn} 
for the corresponding potential symmetry:
$E_v( ((v_t -\b v_t^2)_t - v_{xx})D_t^{-1}P ) =0$. 
This condition can be expressed in terms of $p$
through the potential $v= \partial_t^{-1}p$
and the variational derivative relation
$\delta/\delta v = -D_t \delta/\delta p$.
Thus, the following result holds.

\begin{proposition}
If an infinitesimal symmetry $\hat\X=P\partial_p$ of the dissipative Westervelt equation \eqref{Weqn}
satisfies the condition 
\begin{equation}\label{noether.op.domain}
D_t E_p(( (1 -2\b p)p_t - \partial_t^{-1}p_{xx})D_t^{-1}P ) =0
\end{equation}
off of the solution space $\Esp$, 
then
\begin{equation}\label{Q.P.noether}
Q=\Jop(P) = -(D_t^{-1})^2 P
\end{equation}
is a multiplier that yields a conservation law.
\end{proposition}

This can also be viewed as a nonlocal version of Noether's theorem
by referring to the Lagrangian for the undamped potential equation \eqref{pot2.eqn.undamped}.
First, the Lagrangian implies 
\begin{equation}
E_p(L) = \Jop( (p - \b p^2)_{tt} - p_{xx} )
\end{equation}
where the expression \eqref{pot2.Lagr} for $L$ is nonlocal in terms of $p$:
\begin{equation}
L = \tfrac{1}{2}( (\partial_t^{-1} p_x)^2  -p^2 ) + \tfrac{\b}{3} p^3.
\end{equation}
Next, in terms of this nonlocal Lagrangian,
a variational symmetry can be defined by the condition
\begin{equation}\label{Weqn.varsymm}
0 = E_p( \pr\hat\X(L) ) = E_p( P\Jop( (p - \b p^2)_{tt} - p_{xx} ) )
\end{equation}
which is required to hold off of the solution space $\Esp$.
This condition \eqref{Weqn.varsymm} is readily seen to be
equivalent to the previous condition \eqref{noether.op.domain}
after integration by parts.

To illustrate these results,
consider the Lie point symmetries \eqref{P.symms}.
Substitution of the symmetry characteristic functions
into condition \eqref{Weqn.varsymm}
shows that it is satisfied for $P_1$, $P_2$, and when $\alpha =0$,
$P_3 + \tfrac{\b}{2}P_4$.
Therefore, their span comprises the variational Lie point symmetries of
the nonlocal Lagrangian for the dissipative Westervelt equation.
Through the relations \eqref{rels},
they correspond to multipliers of the potential equation:
\begin{equation}
D_t^{-1}P_1 = -Q^v_{5a} - \tfrac{1}{2\b} Q^v_1,
\quad
D_t^{-1}P_2 = -Q^v_{5b},
\quad
D_t^{-1}(P_3 +\tfrac{\b}{2}P_4) = Q^v_3
\end{equation}
The corresponding conservation laws, as seen in Theorem~\ref{thm:pot2.conslaws},
are nonlocal. 
Specifically, they describe energy \eqref{ener}, momentum \eqref{mom}, and dilation energy \eqref{dilener}, respectively. 
Note that neither $P_3$ nor $P_4$ are variational themselves
and hence they do not yield conservation laws. 

For completeness, it is useful to remark that
$D_t^{-1}$ is properly defined only up to a constant of integration 
which in the present setting of the basic relations \eqref{rels}
is given by a linear combination of the multipliers \eqref{pot2.multrs1}
of the potential equation.
These multipliers belong to the kernel of $D_t$.
Thus,
\begin{equation}\label{invDt}
D_t^{-1}(0) = c_1 + c_2 x
\end{equation}
where $c_1$ and $c_2$ are arbitrary constants.

\subsection{Nonlocal contact symmetries}\label{sec:nonloc.var.symms}

The contact symmetries \eqref{pot2.contactsymm} and \eqref{pot2.contactsymm.funct}
of the potential equation, which are variational,
correspond to nonlocal variational symmetries of the dissipative Westervelt equation:
\begin{equation}\label{contactsymm}
{\hat\X}_\var =
\big( \b v p_x + 2(1-2\b p) v_x + t((2-5\b p) p_x -5\b v_x p_t) +2x( (1-2\b p)^2 p_t -4b v_x p_x) \big)\partial_p
\end{equation}
and
\begin{equation}\label{contactsymm.funct}
{\hat\X}_\var =
\big( f_p p_t +f_{v_x} p_x \big)\partial_p
\end{equation}
where $v=\partial_t^{-1}p$.
Here $f(p,v_x)$ satisfies $f_{pp}=(1-2\b p)f_{v_x v_x}$.

Hierarchies of higher-order variational symmetries will be derived
from a recursion operator later.

\subsection{Transformation to a linear wave equation}\label{sec:linearization}

The conservation law multiplier \eqref{pot2.multr4.funct}
for the undamped potential equation \eqref{pot2.eqn.undamped}
involves a function $f(v_t,v_x)$ of two variables.
This indicates that equation \eqref{pot2.eqn.undamped} can be mapped to
a linear equation by the general method in \Ref{AncBluWol}. 

In outline, the steps go as follows.
First, the variables in multiplier function give the new independent variables, 
$t^*= v_t$ and $x^*= v_x$. 
Because these variables involve derivatives of $v$,
the new dependent variable will be of the form $v^* = w(t,x,v,v_t,v_x)$
such that the mapping will consist of a contact transformation. 
Write
\begin{equation}\label{J}
J= \big|\partial({t^*},{x^*})/\partial(t,x)\big| = v_{tt}v_{xx}-v_{tx}^2
\end{equation}
which is the Jacobian determinant. 
Second, consider the multiplier equation 
\begin{equation}
((v_t -\b v_t^2)_t -v_{xx})F - (F_{v_t v_t} - (1-2\b v_t) F_{v_x v_x})w J
= D_t T + D_x \Phi
\end{equation}
where $F(v_t,v_x)$ is arbitrary function replacing $f(v_t,v_x)$.
Note that the terms involving derivatives of $F$ are of the same form as 
equation \eqref{pot2.f.eqn} which holds for $f$. 
Applying the Euler-Lagrange operator to the multiplier equation
yields a determining equation for $w$: 
\begin{equation}
E_v\big( ((v_t -\b v_t^2)_t -v_{xx})F - (F_{v_t v_t} - (1-2\b v_t) F_{v_x v_x})w J \big) =0.
\end{equation}
Next, this equation splits with respect to derivatives of $F$
and derivatives of $v_t$, $v_x$,
which gives an overdetermined system of linear PDEs for $w(t,x,v,v_t,v_x)$.
The system is straightforward to integrate, yielding 
$w=v-t v_t - x v_x$ up to an arbitrary function of $v_t$ and $v_x$,
which can be put to zero. 
Thus, the new dependent variable is given by $v^* = v-t v_t - x v_x$.

These steps determine the contact transformation:
\begin{equation}\label{linearmapping}
t^*= v_t,
\quad
x^*= v_x,
\quad
v^* = v-t v_t - x v_x, 
\quad
v^*_{t^*} = -t,
\quad
v^*_{x^*} = -x.
\end{equation}
Last, observe that in terms of these new variables,
the linear equation \eqref{pot2.f.eqn} satisfied by the multiplier function $f(v_t,v_x)$ is given by 
\begin{equation}\label{linear.eqn}
v^*_{t^*t^*} = (1-2\b t^*) v^*_{x^* x^*}
\end{equation}
which is self-adjoint.
(Namely, its Frechet derivative, 
$D_{t^*}^2 - (1-2\b t^*) D_{x^*}^2$,
is a self-adjoint operator.)
Note that equation \eqref{linear.eqn} is a wave equation for $t^*<1/(2\b)$
but an elliptic equation for $t^*>1/(2\b)$. 

Through the main theorem in \Ref{AncBluWol},
the potential equation \eqref{pot2.eqn.undamped} for the undamped Westervelt equation 
is mapped into the linear equation \eqref{linear.eqn}
under the contact transformation \eqref{linearmapping}.
This is an instance of a well-known general hodograph (Legendre) transformation that linearizes a class of quasilinear partial differential equations
(see \Ref{ClaFokAbl} and references therein).

\subsection{Recursion operators}\label{sec:recursionop}

The linear equation \eqref{linear.eqn} is manifestly invariant
under translation in $x^*$. 
Hence, it possesses a recursion operator $\Rop^{v^*}=D_{x^*}$,
which maps symmetries in characteristic form
$\hat\X^* = P^*\partial_{v^*}$ into symmetries. 
This structure is inherited by the undamped potential equation \eqref{pot2.eqn.undamped}.
Specifically, under the inverse of the contact transformation \eqref{linearmapping} that maps the potential equation into equation \eqref{linear.eqn}, 
namely 
\begin{equation}\label{inv.linearmapping}
t= -v^*_{t^*},
\quad
x= -v^*_{x^*},
\quad
v = v^* -t^* v^*_{t^*} - x^* v^*_{x^*}, 
\quad
v_t = t^*,
\quad
v_x = x^*, 
\end{equation}
the action of a symmetry yields
\begin{equation}
\hat\X^*(t)= -D_{t^*} P^* =\tau^v,
\quad
\hat\X^*(x)= -D_{x^*} P^* =\xi^v, 
\quad
\hat\X^*(v)= P^* -t^*D_{t^*} P^* -x^* D_{x^*} P^* =\eta^v. 
\end{equation}
The corresponding symmetry of the undamped potential equation is given by
substitution of these expressions into the characteristic function \eqref{Pv.generator},
yielding 
\begin{equation}
P^v = P^* 
\end{equation}
after cancellation of terms. 
This implies that $\Rop^v =\Rop^{v^*}$. 
Then, using $D_{x^*} = J^{-1}(v_{tt}D_x -v_{tx}D_t)$
as obtained through the transformation \eqref{inv.linearmapping},
the recursion operator of equation \eqref{linear.eqn}
is mapped into the operator
\begin{equation}\label{x-transl.recursion.op}
\Rop^v = J^{-1}(v_{tt}D_x -v_{tx}D_t),
\end{equation}
where $J$ is the Jacobian expression \eqref{J}. 

The resulting symmetry recursion operator \eqref{x-transl.recursion.op}
for the undamped potential equation \eqref{pot2.eqn.undamped}
generates a sequence of infinitesimal symmetries
$\hat\X^v_{(k)}=P^v_{(k)} \partial_v$
given by the characteristic functions
\begin{equation}\label{Pv.sequence}
P^v_{(k)} = (\Rop^v)^k P^v,
\quad
k=1,2,\ldots 
\end{equation}
starting from any given symmetry $\hat\X^v =P^v\partial_v$
admitted by the undamped potential equation. 

With $P^v\partial_v$ taken to be the six Lie point symmetries \eqref{pot2.eqn.symms1}--\eqref{pot2.eqn.symms2}, 
the following symmetry characteristic functions are obtained from $\Rop^v$:
\begin{align}
& \Rop^v(P^v_1) = \Rop^v(1)= 0,
\\
& \Rop^v(P^v_2) = \Rop^v(x) = -v_{tt}/J,
\\
& \Rop^v(P^v_3) = \Rop^v(-v_t) = 0, 
\\
& \Rop^v(P^v_4) =\Rop^v(-v_x) = 1,
\\
& \Rop^v(P^v_5) = \Rop^v(t - 2\b t v_t -3\b x v_x) = 3\b x +((1-2\b v_t)v_{tx}+3\b v_x v_{tt})/J,
\\
& \Rop^v(P^v_6) =\Rop^v(v-t v_t -x v_x) = x.
\end{align}

Thus, $\Rop^v$ generates two short sequences
\begin{equation}
P^v_3\to 0
\quad\text{ and }\quad
P^v_4\to P^v_1\to 0
\end{equation}
plus two infinite hierarchies
\begin{equation}\label{R.Pv6.hierarchy}
P^v_6\to P^v_{6,(1)}: =P^v_2 \to P^v_{6,(2)}:= -v_{tt}/J 
\to \cdots
\end{equation}  
and
\begin{equation}\label{R.Pv5'.hierarchy}
P^v_{5'}:= P^v_5-3\b P^v_6 \to P^v_{5',(1)} := ((1-2\b v_t)v_{tx}+3\b v_x v_{tt})/J
\to \cdots
\end{equation}
where
\begin{equation}\label{Pv5'}
P^v_{5'}= t -3\b v + \b t v_t.
\end{equation}

Both hierarchies start from a scaling-type symmetry
which contains $t$ and $x$ explicitly, 
and produce an infinite sequence of higher-order symmetries
represented by the characteristic functions 
\begin{equation}\label{Pv.hierarchies}
P^v_{6,(k+2)} = (\Rop^v)^k P^v_{6,(2)},
\quad
P^v_{5',(k+1)} = (\Rop^v)^k P^v_{5',(1)},
\quad
k=0,1,2,\ldots
\end{equation}
which do not contain $t$ and $x$ explicitly.

Recall that, from the correspondence \eqref{pot2.QP},
neither $P^v_6$ nor $P^v_{5'}$ are variational symmetries,
while $P^v_{6,(1)}=P^v_2$ and the linear combination $P^v_{5'} +\tfrac{7}{2}\b P^v_6$ 
are variational. 
An explicit check of the variational symmetry condition \eqref{pot2.varsymm.deteqn} 
shows similarly that the higher-order symmetries represented by
$P^v_{6,(2)}$, $P^v_{5',(1)}$, $P^v_{5',(2)}$ are not variational,
and both 
\begin{equation}\label{R3.Pv6}
P^v_{6,(3)} =
(v_{tx}^3 v_{ttt} - 3 v_{tt} v_{tx}^2 v_{ttx} + 3 v_{tt}^2 v_{tx} v_{txx} -v_{tt}^3 v_{xxx})/J^3
\end{equation}
and the linear combination
\begin{equation}\label{R2.Pv5'}
\begin{aligned}
P^v_{5',(2)} + \tfrac{1}{2}\b P^v_{6,(2)} = 
& (9\b v_x v_{tt}v_{tx} +z (v_{tx}^2 + 2v_{tt}v_{xx}))(v_{tx} v_{ttx} -v_{tt} v_{txx})/J^3
\\& \quad
+ (3\b v_x v_{tx} +z v_{xx})(v_{tt}^2 v_{xxx} -v_{tx}^2 v_{ttt})/J^3
-\tfrac{7}{2}\b v_{tt}/J
\end{aligned}
\end{equation}
are variational.

The conservation laws arising respectively from
these two variational symmetries \eqref{R3.Pv6} and \eqref{R2.Pv5'}
are given by,
up to local equivalence, 
\begin{equation}\label{conslaw.R3.Pv6}
T  = v_{tx}/J,
\qquad
\Phi=  v_{tt}/J
\end{equation}
after dropping an overall factor of $-\tfrac{1}{2}$, 
and
\begin{equation}\label{conslaw.R2.Pv5'}
T  = ((1-2\b v_t)^2 v_{tt} + 3\b v_x v_{tx})/J,
\qquad
\Phi=  \b x + ((1-2\b v_t)v_{tx} + 3\b v_x v_{tt})/J .
\end{equation}

The apparent pattern here that $P^v_{6,(k)}$ for odd $k$ is variational
and that a linear combination of $P^v_{5',(k)}$ and $P^v_{6,(k)}$ for even $k$ is variational
corresponds to the fact that $D_{x^*}^2$ is a recursion operator
for variational symmetries of the linear equation \eqref{linear.eqn}.
In particular, the resulting two hierarchies of variational symmetries,
both of which do not contain $t$ and $x$ explicitly, 
turn out to be given by
\begin{equation}\label{var.Pv.hierarchy.odd}
P^v_{6,(2l+1)},
\quad
l=1,2,\ldots
\end{equation}
and
\begin{equation}\label{var.Pv.hierarchy.even}
P^v_{5',(2l)} + (\tfrac{7}{2} -3l)\b P^v_{6,(2l)} , 
\quad
l=1,2,\ldots . 
\end{equation}

The recursion operator $\Rop^v$ can also be applied to 
the contact symmetries \eqref{pot2.contactsymm} and \eqref{pot2.contactsymm.funct},
which are variational. 
Their characteristic functions are given by the multipliers \eqref{pot2.contactP}
as shown from the correspondence \eqref{pot2.QP}.
For the latter multiplier $Q^v_5 = P^v_8 =f(v_t,v_x)$,
where this function satisfies equation \eqref{pot2.f.eqn},
the action of the recursion operator is readily seen to amount to replacing
$f(v_t,v_x)$ with $\partial_{v_x} f(v_t,v_x)$.
This action represents an infinitesimal translation symmetry
$\X_{f} = \partial_{v_x}$ with respect to $v_x$.
It thereby maps the family of contact symmetries \eqref{contactsymm.funct},
parameterized by $f(v_t,v_x)$, 
into itself.
Note that the entire family is variational. 

For the multiplier $Q^v_4=P^v_7$ given by expression \eqref{pot2.multr3}, 
the recursion operation yields a hierarchy of infinitesimal symmetries
$\hat\X^v_{(k)} = P^v_{7,(k)}\partial_v$ 
given by the higher-order characteristic functions 
\begin{equation}\label{P.sequence.contact}
P^v_{7,(k)} = (\Rop^v)^k P^v_7,
\quad
k=1,2,\ldots . 
\end{equation}
This hierarchy is independent of
the previous two hierarchies \eqref{R.Pv6.hierarchy} and \eqref{R.Pv5'.hierarchy}.
An explicit check of the variational symmetry condition \eqref{pot2.varsymm.deteqn}
shows that $P^v_{7,(1)}$, $P^v_{7,(2)}$, $P^v_{7,(3)}$ are not variational; 
however,
the linear combinations
$\b P^v_{7,(2)} +2 P^v_{5',(1)}$ and $\b P^v_{7,(4)} +4 P^v_{5',(3)}$
are variational.
The conservation law arising from the first of these variational symmetries
is given by
\begin{equation}\label{conslaw.Pv7}
\begin{aligned}
T = & 
\big( \tfrac{3}{2} \b^2 v_x^2 v_{tx} + \b (1-2\b v_t)^2 v_x v_{tt}
+ \tfrac{1}{6} ((1-2\b v_t)^3 -1)v_{tx} \big)/J ,
\\
\Psi = & 
\big( \tfrac{3}{2} \b^2 v_x^2 v_{tt} + \b (1-2\b v_t)v_x v_{tx}
+ \tfrac{1}{6} ((1-2\b v_t)^3 -1)v_{tt} \big)/J +\b^2 v
\end{aligned}
\end{equation}
up to local equivalence and modulo preceding conservation laws. 
The second conservation law will be omitted because of its length.

From the fact that $D_{x^*}^2$ is a recursion operator
for variational symmetries of the linear equation \eqref{linear.eqn},
the preceding pattern turns out to yield a hierarchy of 
higher-order variational symmetries represented by the characteristic functions
\begin{equation}\label{var.P.contact.hierarchy}
\b P^v_{7,(2l)} +2l P^v_{5',(2l-1)},
\quad
l=1,2,\ldots
\end{equation}
which contain $t$ and $x$ explicitly. 

Another recursion operator
$\Rop^{v^*}_\dil = (2t^* -1/\b)D_{t^*} + 3x^* D_{x^*}$
comes from the manifest invariance of the linear equation \eqref{linear.eqn} 
under dilation of $x^*$ and $t^* - 1/(2\b)$,
namely via the infinitesimal Lie point symmetry
\begin{equation}\label{lineqn.dil.symm}
\X^* = (2t^* -1/\b)\partial_{t^*} + 3x^* \partial_{x^*}.
\end{equation}
The corresponding recursion operator inherited by
the undamped potential equation \eqref{pot2.eqn.undamped} is given by 
\begin{equation}\label{dil.recursion.op}
\Rop^v_\dil = J^{-1}\big( (3 v_x v_{tt} +((1/\b)-2 v_t)v_{tx})D_x -(3 v_x v_{tx} +((1/\b)-2 v_t)v_{xx})D_t \big).
\end{equation}
This operator generates sequences of infinitesimal symmetries
\begin{equation}\label{dil.P.sequence}
\hat\X^v = (\Rop^v_\dil)^k P^v, 
\quad
k=1,2,\ldots
\end{equation}
starting from the Lie point symmetries \eqref{pot2.eqn.symms1}--\eqref{pot2.eqn.symms2}
and the contact symmetries \eqref{pot2.contactsymm}--\eqref{pot2.contactsymm.funct}
in characteristic form \eqref{pot2.contactP}.

Exploration of the properties of these sequences \eqref{dil.P.sequence}
will be considered elsewhere.

\subsection{Higher-order conservation laws and symmetries}\label{sec:higherorder}

The symmetry recursion operators \eqref{x-transl.recursion.op} and \eqref{dil.recursion.op}, 
along with the hierarchies of higher-order symmetries generated by them,
are inherited by the undamped Westervelt equation \eqref{Weqn.undamped}
through the prolongation relations \eqref{rels}.

Prolongation of the recursion operator \eqref{x-transl.recursion.op}
yields 
\begin{equation}\label{Rop}
\Rop = D_t \Rop^v D_t^{-1} = D_t J^{-1}(p_x D_t - p_t D_x) D_t^{-1}
\end{equation}
where this expression is understood to be a composition of operators,
with 
\begin{equation}\label{J.p}
J = p_t v_{xx} -p_x^2
\end{equation}
from expression \eqref{J}. 

Applying the operator $\Rop$ to the Lie point symmetries of the undamped Westervelt equation,
which are given by the characteristic functions \eqref{P.symms},
yields 
$\Rop(P_1) = \Rop(P_2) = \Rop(P_4) = 0$
and 
$\Rop(P_3) = D_t P^v_{5',(1)}$
from the correspondence \eqref{P.Pv} combined with the relation \eqref{P.rel},
where
\begin{equation}\label{P5'.p}
P^v_{5',(1)} = ((1-2\b p)p_x+3\b v_x p_t)/J
\end{equation}
in terms of $p$. 
Furthermore, if the general form \eqref{invDt} for $D_t^{-1}$ is used here,
then $\Rop(0) = c_2 D_t P^v_{6,(2)}$,
where
\begin{equation}\label{P6.p}
P^v_{6,(2)} = -p_t/J
\end{equation}
in terms of $p$. 
The resulting characteristic functions $\Rop(0)$ and $\Rop(P_3)$
have the explicit form
\begin{equation}\label{RP6u}
P_{(1)} := D_t P^v_{6,(2)}
= \big( p_x^2p_{tt} - 2 p_t p_x p_{tx} + p_t^2 p_{xx} \big)/J^2
\end{equation}
and 
\begin{equation}\label{RP5'u}
\begin{aligned}
P_{(1)'} := D_t P^v_{5',(1)}
& = \big( (1-2\b p)( v_{xx}(p_t p_{tx} - p_x p_{tt}) + p_x(p_x p_{tx} -p_t p_{xx}) ) 
\\&\qquad
-3 v_x( p_x^2p_{tt} - 2 p_t p_x p_{tx} + p_t^2 p_{xx}) \big)/J^2 
+\b p_tp_x/J . 
\end{aligned}
\end{equation}
which represent second-order symmetries 
of the undamped Westervelt equation \eqref{Weqn.undamped}. 
They are respectively the root symmetries in the hierarchies given by 
\begin{equation}\label{P.hierarchies}
P_{(k)} = (\Rop)^{k-1} P_{(1)},
\quad
P_{(k)'} = (\Rop)^{k-1} P_{(1)'},
\quad
k=1,2,\ldots
\end{equation}
which correspond to the two hierarchies \eqref{Pv.hierarchies}
admitted by the undamped potential equation.

An additional hierarchy of higher-order symmetries
is inherited through the hierarchy \eqref{P.sequence.contact}
starting from the infinitesimal contact symmetry \eqref{contactsymm} 
of the undamped potential equation.
This root symmetry is given by the characteristic function 
\begin{equation}\label{Q4u}
P_{(1)''} := D_t P^v_7 =
\b p_x v + (2(1-2\b p) -5\b t p_t - 8\b x p_x)v_x
+ t (2 -5 \b p)p_x + 2 x(1-2\b p)^2 p_t
\end{equation}
which represents a first-order symmetry
of the undamped Westervelt equation \eqref{Weqn.undamped}. 
In the resulting hierarchy 
\begin{equation}\label{P.contact.hierarchy}
P_{(k)''} := (\Rop)^{k-1} P_{(1)''},
\quad
k=1,2,\ldots, 
\end{equation}
all of the symmetries involve $t$ and $x$ explicitly,
in contrast to the symmetries in the two hierarchies \eqref{P.hierarchies}.

On solutions $p(t,x)$ of equation \eqref{Weqn.undamped},
note that 
\begin{equation}
J|_\Esp = (1-2\b p) p_t^2 -p_x^2 := J_\Esp
\end{equation}
and $v_{xx}|_\Esp = (1-2\b p)p_t$
are local in terms of $p$.
Hence,
$P_{(1)}|_\Esp$ is local,
whereas $P_{(1)'}|_\Esp$ and $P_{(1)''}|_\Esp$ are nonlocal
due to the presence of $v_x$ and $v$.
The same feature holds for the higher-order symmetry characteristics
in the respective three hierarchies \eqref{P.hierarchies} and \eqref{P.contact.hierarchy}.
Specifically, this feature can be seen via the relations
\begin{equation}
(\Rop)^k|_\Esp = D_t (\Rop^v)^k|_\Esp D_t^{-1},
\quad
\Rop^v|_\Esp = J^{-1}_\Esp (p_xD_t -p_tD_x)
\end{equation}
showing that the operator $(\Rop)^k|_\Esp D_t$ is local in terms of $p$,
while expressions \eqref{P5'.p}, \eqref{P6.p}, \eqref{pot2.multr3}
respectively show that 
$(D_t^{-1} P_{(1)})|_\Esp = -p_t/J_\Esp$ is local,
and both
$(D_t^{-1} P_{(1)'})|_\Esp = ((1-2\b p)p_x+3\b v_x p_t)/J_\Esp$
and 
$(D_t^{-1} P_{(1)'''})|_\Esp = v v_x + t (\tfrac{2}{\b}  - 5 p)v_x -x (4v_x^2 +\tfrac{1}{3\b^2}(1-2\b p)^3)$
are nonlocal. 

The variational symmetry condition \eqref{Weqn.varsymm} determines
which of the higher-order infinitesimal symmetries
in the three hierarchies \eqref{P.hierarchies} and \eqref{P.contact.hierarchy}
correspond to multipliers for conservation laws of
the undamped Westervelt equation \eqref{Weqn.undamped}. 
This condition can be checked explicitly for any given infinitesimal symmetry
or linear combination of given infinitesimal symmetries.
(Note that the variational property does not persist in general 
when a symmetry characteristic is evaluated on solutions.)
More simply,
all variational symmetries can be found through the relation \eqref{rels}
by prolongation of the variational symmetries of the undamped potential equation \eqref{pot2.eqn.undamped}. 
This yields three hierarchies of higher-order variational symmetries
represented by the characteristic functions 
\begin{align}
& P_{\var (l)} = P_{(2l)} = D_t(P^v_{6,(2l+1)}),
\quad
l=1,2,\ldots
\label{Pv6.hierarchy}
\\
& P_{\var (l)'}= P_{(2l)'} + (\tfrac{7}{2} -3l)\b P_{(2l-1)}
= D_t(P^v_{5',(2l)} + (\tfrac{7}{2} -3l)\b P^v_{6,(2l)}),
\quad
l=1,2,\ldots
\label{Pv5'.hierarchy}\\
& P_{\var (l)''} = \b P_{(2l+1)''} +2l P_{(2l-1)'}
= D_t(\b P^v_{7,(2l)} +2l P^v_{5',(2l-1)}), 
\quad
l=1,2,\ldots
\label{Pv7.hierarchy}
\end{align}
which are inherited respectively from the three hierarchies
\eqref{var.Pv.hierarchy.odd}, \eqref{var.Pv.hierarchy.even}, \eqref{var.P.contact.hierarchy}
for equation \eqref{pot2.eqn.undamped}.  
The Noether operator \eqref{Noether.op} then yields 
the multiplier expression \eqref{Q.P.noether} 
from each variational symmetry in these hierarchies. 

A third way of obtaining the multipliers for higher-order
conservation laws of the undamped Westervelt equation \eqref{Weqn.undamped}
is by use of the adjoint-symmetry recursion operator 
\begin{equation}\label{multr.Rop}
\Rop_Q = \Jop\Rop\Jop^{-1} =  D_t^{-1}J^{-1}(p_x D_t - p_t D_x)D_t
\end{equation}
which arises from composing the symmetry recursion operator \eqref{Rop}
with the Noether operator \eqref{Noether.op}. 
Note that $\Rop_Q = D_t^{-1}\Rop^v D_t$ holds in accordance with the relations \eqref{rels},
where these expressions are understood to be a composition of operators. 

Applying $\Rop_Q$ to the four lowest-order multipliers \eqref{multr.loword}
of the undamped Westervelt equation 
yields
$\Rop_Q(Q_1) = \Rop_Q(Q_2) = \Rop_Q(Q_3) =0$,
which are trivial;
and $\Rop_Q(Q_4) = D_t^{-1}R^v(x) = D_t^{-1}(-p_t/J)$
which thereby gives 
\begin{equation}
\Rop_Q(Q_4) = D_t^{-1}P^v_{6,(2)} = -\Jop(P_{(1)})
:= Q_{(1)}
\end{equation}
from expressions \eqref{P6.p}--\eqref{RP6u}.
Note that $Q_{(1)}$ is an adjoint-symmetry but is not a multiplier
since $P^v_{6,(2)}$ is non-variational.
Continuing, it is simple to see 
\begin{equation}
\Rop_Q^2(Q_4) = D_t^{-1}P^v_{6,(3)} = -\Jop(P_{(2)}) = -\Jop(P_{\var(1)}) 
:=Q_{(2)} , 
\end{equation}
which is a multiplier,
since $P^v_{6,(3)}$ is variational.

It is useful to note that 
if the general form \eqref{invDt} for $D_t^{-1}$ is used in $\Rop_Q$, 
then $D_t^{-1}(0) = c_1 Q_1 + c_2 Q_2$
and $D_t^{-2}(0) = D_t^{-1}(c_1+c_2 x)= c_1 Q_3 + c_2 Q_4$,
whereby 
\begin{equation}
Q_4 = D_t^{-1}P^v_2 = D_t^{-1}P^v_{6,(1)} :=Q_{(0)} .
\end{equation}
Thus, when the operator
\begin{equation}
\Rop_Q^2 =\Jop\Rop^2\Jop^{-1} =  D_t^{-1}(J^{-1}(p_x D_t - p_t D_x))^2D_t
\end{equation}
is applied to the multiplier $Q_4$,
this generates a sequence of multipliers
\begin{equation}\label{Q4.hierarchy}
\begin{aligned}
& Q_{(0)} = D_t^{-1}P^v_{6,(1)}= -\Jop(0)|_{c_1=0,c_2=1} 
\to Q_{(2)} = D_t^{-1}P^v_{6,(3)} = -\Jop(P_{\var(1)}) \\
& \to Q_{(4)} := D_t^{-1}P^v_{6,(5)} = -\Jop(P_{\var(2)}) \to \cdots ,
\end{aligned}
\end{equation}
which corresponds to the hierarchy of variational symmetries \eqref{Pv6.hierarchy}
through the Noether operator \eqref{Noether.op}. 

In addition, if $v_{xx}$, $v_{xxx}$, and $v_{txx}$ in $P^v_{6,(3)}$
are expressed in terms of $v_{tt}$, $v_{ttx}$, and $v_{ttt}$ 
through the undamped potential equation \eqref{pot2.eqn.undamped},
then it is straightforward to obtain the relation
\begin{equation}
P^v_{6,(3)}|_{\Esp^v} = D_t Q_5
\end{equation}
where $Q_5$ is the multiplier \eqref{multr.loword.ais0}.
Consequently,
$Q_{(2)}|_{\Esp^v} = \Rop_Q^2(Q_4)|_{\Esp^v} =Q_5$. 
In a similar way, applying $\Rop_Q^2$ to $Q_5$ leads to the relation
$Q_{(4)}|_{\Esp^v} = \Rop_Q^2(Q_5)|_{\Esp^v}$.
Hence, the resulting hierarchy of higher-order multipliers
generated by $\Rop_Q$ applied to $Q_5$
in essence belongs to the same hierarchy \eqref{Q4.hierarchy}
that arises from $\Rop_Q$ applied to $Q_4=Q_{(0)}$. 

The two lowest-order conservation laws arising from
the hierarchy \eqref{Q4.hierarchy} of multipliers 
are respectively given by conserved current \eqref{pot2.conslaw2} with $\alpha=0$
(which is locally equivalent to conserved current \eqref{conslaw4} with $\alpha=0$) 
and conserved current \eqref{conslaw5}
(which is locally equivalent to the conservation law \eqref{conslaw.R3.Pv6}).
These two conservation laws are local in terms of $p$.
The next conservation law, which arises from $Q_{(4)}$,
is also local in terms of $p$:
\begin{subequations}
\begin{align}
&\begin{aligned}
T = &
\tfrac{1}{2} (\z p_{tt}^2 +p_{tx}^2) (5 \z^2 p_t^4 + 10 \z p_t^2 p_x^2 + p_x^4)p_x/J_\Esp^5
- \yone\ytwo_+ (\z^2 p_t^4 + 10 \z p_t^2 p_x^2 + 5 p_x^4) p_t/(\z J_\Esp^5)
\\&\quad
- 10 \b \yone (\z p_t^2 + p_x^2) p_t^2 p_x/(\z J_\Esp^4) - 4\b \yone p_x /(\z^4 p_t^4)
- 8\b \ytwo_- p_x^2 /(\z^5 p_t^5)
\\&\quad
- 10 \b^2 (\z p_t^2 + 3 p_x^2) p_t^6 p_x/J_\Esp^5 ,
\end{aligned}
\label{T6}
\\
&\begin{aligned}
\Phi = &
-p_{tt} p_{tx}(5 \z^2 p_t^4 + 10 \z p_t^2 p_x^2 + p_x^4) p_x/J_\Esp^5
+ ( \z \yone^2 + \ytwo_+^2 )(\z^2 p_t^4 + 10 \z p_t^2 p_x^2 + 5 p_x^4) p_t/(2 \z^2 J_\Esp^5)
\\&\quad
+10\b p_{tx} (\z p_t^2 + p_x^2)p_t^2p_x/(\z J_\Esp^4)
- 4\b p_{tx} p_x/(\z^4 p_t^4) 
+ 8 \b p_{tt} p_x^2/(\z^4 p_t^5)
\\&\quad
- 16 \b^2 p_x^2/(\z^5p_t^3) - 10 \b^2 p_t^3 p_x^4 (3 \z p_t^2 + p_x^2)/(\z^2 J_\Esp^5) ,
\end{aligned}
\label{Phi6}
\end{align}
\end{subequations}
where
$\z=1-2\b p$,
$\yone =\z p_{tt} -2\b p_t^2$, $\ytwo_\pm = \z p_{tx} \pm 2\b p_t p_x$. 

The resulting higher-order conserved integrals 
$C_{(k)} = \int_{-\infty}^{\infty} T_{(k)}\,dx$, $k=1,2,\ldots$,
which correspond to the multipliers $Q_{(2k)} = -\Jop(P_{\var (k)})$, 
exhibit the following features
as seen explicitly for $C_{(1)} = C_5$ given by the integral \eqref{C5},
and $C_{(2)}$ given by the integral of the density \eqref{T6}. 
$C_{(k)}$ has order $k$ in terms of $p$;
$T_{(k)}$ is local and has odd parity under reflection $(t,x)\to (-t,-x)$
on derivatives of $p$; 
the highest power of $J_\Esp$ in the denominator of $T_{(k)}$ is $4k-3$.
This parity property is analogous to the odd spatial parity of
the nonlocal momentum \eqref{mom} and dilation momentum \eqref{dilmom}. 

In a similar way, the two other hierarchies 
of variational symmetries \eqref{Pv5'.hierarchy} and \eqref{Pv7.hierarchy}
yield multipliers defined by the Noether correspondence \eqref{noether.rel}.
The resulting conserved integrals 
are nonlocal in terms of $p$ but share the other features of
the conserved integrals $C_{(k)}$. 
In particular, the lowest-order one in each hierarchy is given by
\begin{equation}
C_{(1)'} = \int_{-\infty}^{\infty} ((1-2\b p)^2 p_t + 3\b v_x p_x)/J_\Esp \,dx
\end{equation}
and
\begin{equation}
C_{(1)''} = \int_{-\infty}^{\infty} \big(
\tfrac{3}{2} \b^2 v_x^2 p_x + \b (1-2\b p)^2 v_x p_t
+ \tfrac{1}{6} ((1-2\b p)^3 -1)p_x
\big)/J_\Esp \,dx
\end{equation}
from conservation laws \eqref{conslaw.R2.Pv5'} and \eqref{conslaw.Pv7} respectively,
which contain $v_x = \partial_t^{-1} p_x$. 
Note that this nonlocal variable has even parity
under reflection $(t,x)\to (-t,-x)$ acting on derivatives of $p$. 

As a final observation,
note that the nonlocal energy \eqref{ener}, momentum \eqref{mom},
dilational energy \eqref{dilener} and dilational momentum \eqref{dilmom}
are not part of the preceding three hierarchies of conserved integrals.
Instead they belong to the family of generalized energy-momentum integrals \eqref{gen.enermom}.

\section{Concluding remarks}\label{sec:conclude}

The dissipative Westervelt equation \eqref{Weqn} is found to possess
six conservation laws of low-order, 
four of which are local while the other two are nonlocal. 
These conservation laws describe conserved quantities 
related to the net mass and weighted net mass displaced by sound waves. 

In the undamped case, equation \eqref{Weqn} is further found to possess 
three nonlocal conservation laws of first order. 
One of these conservation laws describes a family of generalized energy-momentum quantities 
which includes energy and momentum as special cases. 
The other two conservation laws describe dilational energy and dilational momentum. 

As a main result, it is shown that the undamped Westervelt equation 
can be mapped into a linear equation by a contact (hodograph) transformation applied to a potential system
which has a Lagrangian formulation. 
Through this mapping, the undamped equation inherits
several ``hidden'' (nonlocal) variational structures:
a Lagrangian, a Hamiltonian, and a Noether operator. 
Additionally, the Lie point symmetries of the linear equation
give rise to recursion operators 
and associated hierarchies of symmetries and conservation laws,
which are inherited by the undamped equation. 
It is remarkable that one of the inherited hierarchies turns out to be local,
namely, consisting of higher-order local symmetries
and higher-order local conservation laws.

Several aspects can be pursued further: 
exploration of more features of the hierarchies of symmetries and conservation laws 
and their physical meaning; 
use of the symmetries to derive exact solutions of
both the dissipative and undamped Westervelt equation;
use of the conservation laws to study analysis of solutions. 

All of the results obtained in the present paper illustrate the numerous useful 
applications of modern symmetry analysis for studying nonlinear wave equations.

\section*{Acknowledgments}
SCA is supported by an NSERC Discovery Grant. 
APM, TMG, and MLG acknowledge the support from the research group FQM-201 of the \textit{Junta de Andaluc\'ia}.

\section*{Appendix A: Jet space}\label{app.A}

The jet space of a variable $w(t,x)$ is the coordinate space $(t,x,w,\partial w,\partial^2 w,\ldots)$ 
where 
$\partial=(\partial_t,\partial_x)$, $\partial^2=(\partial_t^2,\partial_t\partial_x,\partial_x^2)$, and so on. 
Total derivatives are denoted by
\begin{equation}
\begin{aligned}
D_t & =\partial_t + w_t\partial_w + w_{tt}\partial_{w_t} + w_{tx}\partial_{w_x} +\cdots
\\
D_x & =\partial_x + w_x\partial_w + w_{tx}\partial_{w_t} + w_{xx}\partial_{w_x} +\cdots
\end{aligned}
\end{equation}
They satisfy the property $D w=\partial w$. 

The Euler operator (variational derivative) with respect to $w$ is given by
\begin{equation}
E_w = \partial_w +(-D)\partial_{\partial w} + (D^2)\partial_{\partial^2 w} +\cdots
\end{equation}
where
$D=(D_t,D_x)$, $D^2=(D_t^2,D_t D_x,D_x^2)$, and so on. 
It has the property that $E_w(f)=0$ iff
$f=D_t F^t + D_x F^x$ where $(F^t,F^x)$ are functions in jet space.

\section*{Appendix B: Computation of symmetries and multipliers}\label{app.B}

The classification of infinitesimal Lie point symmetries 
in Theorems~\ref{thm:pointsymms} and~\ref{thm:pot.pointsymms}
is obtained by solving the respective determining equations
\eqref{symm.deteqn} and \eqref{pot2.eqn.symm.deteqns}.
There are four main steps,
which will be explained for Theorem~\ref{thm:pointsymms}.
The proof of Theorem~\ref{thm:pot.pointsymms} is similar.

First,
expression \eqref{P.generator} containing the unknowns
$\eta(t,x,p)$, $\tau(t,x,p)$, $\xi(t,x,p)$ 
is substituted into equation \eqref{symm.deteqn},
and then $p_{xx}$ and its derivatives are substituted via 
the dissipative Westervelt equation \eqref{Weqn},
which carries out the evaluation on $\Esp$;
second, the resulting equation is split with respect to all derivatives of $p$,
which yields an overdetermined system of 12 linear PDEs
on $\eta(t,x,p)$, $\tau(t,x,p)$, $\xi(t,x,p)$,
containing $\alpha$ and $\beta\neq 0$.
Third, the Maple command `rifsimp' is used to find all cases
for which this system is reduced to an involutive form \cite{Sei-book}. 
This is essentially a nonlinear problem because
the parameters $\alpha$ and $\beta$ must be treated as unknowns
and they appear in products with $\eta(t,x,p)$, $\tau(t,x,p)$, $\xi(t,x,p)$;
involutivity leads to a case split: $\alpha=0$ and $\alpha\neq0$.
The resulting involutive systems have a triangular form,
consisting of 9 linear PDEs in each case. 
Last, these systems are readily integrated to get the general solution,
which completes the proof.

Likewise, the classification of low-order multipliers 
in Propositions~\ref{prop:multr.loword} and~\ref{prop:pot.multr.loword}
is obtained by solving the respective determining equations
\eqref{multr.deteqn} and \eqref{pot2.multr.deteqn}. 
The main steps will be explained for proving Proposition~\ref{prop:multr.loword}; the proof of Proposition~\ref{prop:pot.multr.loword} is similar. 

First, the classification is divided into the cases $\alpha=0$ and $\alpha\neq0$.
In the first case, the unknown $Q(t,x,p,p_t,p_x)$
is substituted into equation \eqref{multr.deteqn}
which splits with respect to all second-order derivatives of $p$.
This yields an overdetermined system of 5 linear PDEs
on $Q(t,x,p,p_t,p_x)$, as well as $\beta\neq 0$.
Then the Maple command `rifsimp' is used to find all cases
for which this system is reduced to an involutive form \cite{Sei-book}. 
No case splitting arises, and the resulting system consists of 7 linear PDEs. 
Four of these PDEs are single-term equations that are easily integrated,
while the remaining three PDEs then become two-term equations which 
are straightforward to integrate,
yielding the general solution of the reduced system. 

In the second case, the unknown is $Q(t,x,p,p_t,p_x,p_{tt},p_{tx},p_{xx})$.
Substitution into equation \eqref{pot2.multr.deteqn},
followed by splitting with respect to all third-order and fourth-order
derivatives of $p$,
yields an overdetermined system of 6 linear PDEs. 
Use of the Maple command `rifsimp' shows that the system reduces to an involutive form without case splitting. 
This reduced system consists of 8 linear PDEs,
each of which is a single-term equation.
Direct integration yields the general solution of the system.

\end{document}